\documentclass{aa}

\newcommand{\totaln}{493}
\newcommand{\intn}{97}
\newcommand{\intnhalfhalf}{24}
\newcommand{\srctot}{50}
\newcommand{\srcscal}{11}
\newcommand{\perchadr}{56.2}
\newcommand{\totalnf}{3634}
\newcommand{\srcscalf}{712}
\newcommand{\totalf}{178}
\newcommand{\perchadrf}{7.9}

\usepackage[utf8]{inputenc} 
\usepackage{natbib}
\bibpunct{(}{)}{;}{a}{}{,}
\graphicspath{{figures/}}
\usepackage{amsmath,amssymb}
\usepackage[varg]{txfonts}
\usepackage{wasysym}
\usepackage{marvosym}
\usepackage{textcomp}

\usepackage{graphicx}
\usepackage{xspace}
\usepackage[dvipsnames]{xcolor}

\newcommand{\Swift}{\textsl{Swift}\xspace}

\newcommand{\Fermi}{\textsl{Fermi}\xspace}

\newcommand{\ROSAT}{\textsl{ROSAT}\xspace}

\usepackage{pdflscape}
\makeatletter
\ifcase \@ptsize \relax
  \newcommand{\miniscule}{\@setfontsize\miniscule{7}{8}}
\or
  \newcommand{\miniscule}{\@setfontsize\miniscule{6}{7}}
\or
  \newcommand{\miniscule}{\@setfontsize\miniscule{6}{7}}
\fi
\makeatother

\makeatletter
\newcommand{\mathleft}{\@fleqntrue\@mathmargin\parindent}
\newcommand{\mathcenter}{\@fleqnfalse}
\makeatother

\title{\Fermi-LAT counterparts of IceCube neutrinos above 100\,TeV}
\author{
  F. Krau\ss{}\inst{\ref{affil:api},\ref{affil:grappa}}
  \and K.~Deoskar\inst{\ref{affil:iit},\ref{affil:stock},\ref{affil:remeis}}
  \and C. Baxter\inst{\ref{affil:api},\ref{affil:remeis}}
  \and M.~Kadler\inst{\ref{affil:wuerzburg}}
  \and M.~Kreter\inst{\ref{affil:za},\ref{affil:wuerzburg}}
  \and M.~Langejahn\inst{\ref{affil:wuerzburg}}
  \and K.~Mannheim\inst{\ref{affil:wuerzburg}}
  \and P.~Polko\inst{{\ref{affil:losa}}}
  \and B. Wang\inst{\ref{affil:jh},\ref{affil:remeis}}
  \and J.~Wilms\inst{\ref{affil:remeis}}
}

\institute{
  Anton Pannekoek Institute for Astronomy, University of
  Amsterdam,  Science Park 904, 1098 XH Amsterdam, The Netherlands\\
  \email{fe@fekrauss.com}
  \label{affil:api}
  \and
  GRAPPA, University of Amsterdam,  Science Park 904, 1098 XH Amsterdam, The Netherlands
  \label{affil:grappa}
  \and
  Department of Physics, Indian Institute of Technology Kharagpur,
  West Bengal 721302, India 
  \label{affil:iit}
  \and
  Oskar Klein Centre and Dept. of Physics, Stockholm University,
  SE-10691 Stockholm, Sweden
  \label{affil:stock}
  \and
  Dr.~Remeis Sternwarte \& ECAP, Universit\"at Erlangen-N\"urnberg,
  Sternwartstrasse 7, 96049 Bamberg, Germany
  \label{affil:remeis}
  \and
  Institut f\"ur Theoretische Physik und Astrophysik, Universit\"at
  W\"urzburg, Emil-Fischer-Str.\ 31, 97074 W\"urzburg, Germany
  \label{affil:wuerzburg}
  \and
  Centre for Space Research, North-West University, Private Bag X6001,
  Potchefstroom 2520, South Africa
  \label{affil:za}
  \and
  Theoretical Astrophysics, T-2, MS B227, Los Alamos National Laboratory,
  Los Alamos, NM 87545, USA
  \label{affil:losa}
  \and
  Department of Physics and Astronomy, Johns Hopkins University,
  Baltimore, Maryland 21218, USA
  \label{affil:jh}
}

\authorrunning{F. Krau{\ss}~et al.}
\titlerunning{3LAC counterparts of IceCube neutrinos above 100\,TeV}
\date{Received $<$date$>$ / Accepted $<$date$>$}

\abstract
{
  The IceCube Collaboration has published four years of data and the
  observed neutrino flux is significantly in excess of the 
  expected atmospheric background. Due to the steeply falling
  atmospheric background spectrum, events at the highest energies are
  most likely extraterrestrial. In our previous approach we have
  studied blazars as the possible origin of the High-Energy Starting
  Events (HESE) neutrino events at PeV energies. In this work we
  extend our study to include all HESE neutrinos (which does not
  include IC\,170922A) at or above a
  reconstructed energy of 100\,TeV, but below 1\,PeV. We study the
  X-ray and $\gamma$-ray data of all ($\sim200$)
  3LAC blazars that are positionally consistent with the neutrino
  events above 100\,TeV to determine the maximum neutrino flux from
  these sources. This larger sample allows us to better constrain the
  scaling factor between the observed and maximum number of neutrino
  events. We find that when we consider a realistic neutrino spectrum
  and other factors, the number of neutrinos is in good agreement with
  the detected number of IceCube HESE events. We also show that there
  is no direct correlation between \Fermi-LAT $\gamma$-ray flux and
  the IceCube neutrino flux and that the expected number of neutrinos
  is consistent with the non-detection of individual bright blazars.
}


\begin{document}

\mathcenter
\maketitle

\section{Introduction}\label{sec-intro}
The detection of high-energy PeV neutrinos in the first three years of
operations of the IceCube detector
\citep{Icecube2013a,Icecube2013b,IC3} has led to a great effort to
find the astrophysical origin of these neutrinos. Proposed
counterparts include Sagittarius A$^\ast$, (choked-jet) 
gamma-ray bursts (GRBs), active galactic nuclei (AGN), fast radio
bursts (FRBs), star formation galaxies, magnetar flares, accretion
disks, ultraluminous X-ray sources, and tidal disruption events
\citep{Tamborra2016,Bednarek2016,Bechtol2017,Bustamante2017,Fujita2017,Guepin2017,Aartsen2017,Albert2017,Fahey2017}.
Among these candidate sources, blazars have found particular interest
\citep{eb,Kadler2016,txs} However, several studies have found that blazars
alone are unlikely counterparts because they have only a small
contribution to the
neutrino spectrum for all of the IceCube neutrinos
\citep{Turley2016,Murase2016,Aartsen2016,Feyereisen2017,Ando2017,Neronov2017,Zhang2017}.
Constraints from multiplets indicate that BL Lacs do not contribute
strongly to the observed neutrinos, but the same is not true for FSRQs
\citep{Murase2018}. 
Moreover, the $\gamma$-ray brightest AGN sky show only a weak
positional agreement with HESE neutrinos, demonstrating that the >100
TeV IceCube signal is not simply dominated by a small number of the
$\gamma$-ray brightest blazars. Instead, a larger number of sources
have to contribute to the signal, with each individual source having
only a low Poisson probability for producing an event in multi-year
integrations of current neutrino detectors \citep{Aartsen2015,tannew}.

Owing to the steeply falling atmospheric background with increasing
energy, PeV events are highly likely of extraterrestrial origin. For
this reason, we have concentrated in our previous works on such events
at energies above 1 PeV \citep{eb,Kadler2016}.
Between 100 TeV and 1 PeV, the HESE analysis of the IceCube team has
found 16 additional events \citep{IC3}, whose individual 'signalness'
is reduced, however, so that only about 14 out of these 16
events can be expected to be of astrophysical origin \citep{IC3,IC4yr}.

In addition to the HESE analysis, IceCube has published six Extremely
High Energy (EHE) events via GCN\footnote{available at
  \url{https://gcn.gsfc.nasa.gov/gcn3_archive.html}. GCN numbers
  available for 5 EHE events: 19787, 20247, 20929, 21916, 22105. The
  sixth event is online at
  \url{https://gcn.gsfc.nasa.gov/notices_amon/6888376_128290.amon}.}.
One of these EHE event alerts has recently prompted an intriguing
coincidence of a track-like IceCube neutrino event above 100 TeV, IC
170922A (GCN \#21916), with a blazar outburst \citep{atel,txs,txs2}.
While the signalness of the recent event is only $\sim56\%$, this new
coincidence is reminiscent of another blazar outburst found at the
same time and position as a 2\,PeV neutrino event \citep{Kadler2016},
but due to the track-like nature of the event, the chance coincidence
is substantially lower, on the order of $3 \sigma$. If one $> 100$\,TeV 
neutrino is indeed associated with a blazar, it appears
straightforward to test whether viable blazar candidates exist in the
fields of the 16 published HESE neutrino events between 100\,TeV and
1\,PeV. In this paper, we study possible blazar counterparts to the
four-year IceCube HESE neutrinos above 100\,TeV and below 1\,PeV. We
collect multiwavelength data from X-ray to high-energy $\gamma$ rays
and describe the high-energy peak in the spectral energy distribution (SED) with a logarithmic
parabola. From the integrated energy flux we calculate the maximum
number of neutrinos. This large sample of neutrinos and possible
candidate sources allows us to constrain a scaling factor between the
observed number of neutrinos and the maximum number of neutrinos
calculated from the energy flux. This was previously constrained and
explained by \citet{Kadler2016}, but was only based on the sky area of
one neutrino event ($\sim 800$\,deg$^2$), and referred only to PeV neutrino
events. We further study if the non-detection of neutrinos from
some bright blazars and the small number of source candidates for
neutrinos with small angular uncertainties are still consistent with
blazars as the origin of IceCube PeV neutrinos.

In Sect.\,2 we explain the methods, including the data analysis and
SED generation. In Sect.\,3 we show the results, discussing the
expected number of neutrinos and how we scale that number. Sect.\,4
lists caveats of the methods and how they affect the results. In the
final section we discuss the results and implications for hadronic
blazars. Neutrino event numbers and SEDs are shown in the appendix.

\section{Methods}
\subsection{IceCube HESE events}
In Table~\ref{tab-nus}, we give the full list of HESE events above
100\,TeV in four years of IceCube data \citep{IC4yr}. Sources above
1\,PeV are listed, but not considered in this study, as we have
investigated them before \citep{eb,Kadler2016}.

\renewcommand*{\arraystretch}{1.3}
\begin{table*}\centering
  \caption {IceCube four-year HESE events above 100\,TeV. Events above
    1\,PeV are marked in gray, as they are not considered in this
    study. The first column gives the IceCube event number, the second
  column gives the energy deposited within the IceCube detector, and the
  third column gives the MJD of the event detection. Columns 4 and
  5 give the best-fit right ascension and declination of the
  neutrino event, while the sixth column shows the angular uncertainty
  of the event, and the last column gives the event morphology.}
  \label{tab-nus}
  \begin{tabular}{lllllll}
    IC & $E_\mathrm{deposited}$ & MJD &
    $\alpha_{\mathrm{J2000.0}}$ [$^\circ$] & $\delta_\mathrm{J2000.0}$
    [$^\circ$] & ang. & morphology\\
    & [TeV] & & & & res. & \\
    \hline
2 & $117_{-14.6}^{15.4}$ & 55351.4659661        &       282.6   &       -28     &       25.4    &        Shower  \\
4 & 165.4$_{-14.9}^{19.8}$ &    55477.3930984   &       169.5   &       -51.2   &       7.1     &        Shower  \\
12 & 104.1$_{-13.2}^{12.5}$ &   55739.4411232   &       296.1   &       -52.8   &       9.8     &        Shower  \\
13 & 252.7$_{-21.6}^{25.9}$ &   55756.1129844   &       67.9    &       40.3    &          <1.2  &        Track  \\
\color{gray}14 & \color{gray}1040.7$_{-144.4}^{131.6}$ &        \color{gray}55782.5161911       &       \color{gray}265.6       &\color{gray}   -27.9   &\color{gray}   13.2    &\color{gray}     Shower \\
17 & 199.7$_{-26.8}^{27.2}$ &   55800.3755483   &       247.4   &       14.5    &       11.6    &        Shower  \\
\color{gray}20 &\color{gray} 1140.8$_{-132.8}^{142.8}$ &\color{gray}    55929.3986279   &\color{gray}   38.3    &\color{gray}   -67.2   &       \color{gray}10.7        &\color{gray}     Shower \\
22 & 219.5$_{-24.4}^{21.2}$ &   55941.9757813   &       293.7   &       -22.1   &       12.1    &        Shower  \\
26 & 210.0$_{-25.8}^{29.0}$ &  55979.2551750  & 143.4   &       22.7    &       11.8    &        Shower  \\
30 & 128.7$_{-12.5}^{13.8}$ &   56115.7283574   &       103.2   &       -82.7   &       8       &        Shower  \\
33 & 384.7$_{-48.6}^{46.4}$ &   56221.3424023   &       292.5   &       7.8     &       13.5    &        Shower  \\
\color{gray}35 &\color{gray} 2003.7$_{-261.5}^{236.2}$ &\color{gray}    56265.1338677   &\color{gray}   208.4   &\color{gray}   -55.8   &\color{gray}   15.9    &\color{gray}     Shower \\
38 & 200.5$_{-16.4}^{16.4}$ &   56470.1103795   &       93.3    &       14      &          <1.2  &        Track  \\
39 & 101.3$_{-11.6}^{13.3}$ &   56480.6617877   &       106.2   &       -17.9   &       14.2    &        Shower  \\
40 & 157.3$_{-16.7}^{15.9}$ &   56501.1641008   &       143.9   &       -48.5   &       11.7    &        Shower  \\
45 & 429.9$_{-49.1}^{57.4}$ &   56679.2044683   &       219     &       -86.3   &          <1.2  &        Track  \\
46 & 158.0$_{-16.6}^{15.3}$ &   56688.0702948   &       150.5   &       -22.3   &       7.6     &        Shower  \\
48 & 104.7$_{-10.2}^{13.5}$ &   56705.9419933   &       213     &       -33.2   &       8.1     &        Shower  \\
52 & 158.1$_{-18.4}^{16.3}$ &   56763.5448147   &       252.8   &       -54     &       7.8     &        Shower  \\
\end{tabular}
\end{table*}

\subsection{X-ray data}
The  maximum possible neutrino flux of a blazar can be derived from
the integrated X-ray to $\gamma$-ray high-energy emission hump
\citep[see][and Sect.~\ref{sect-maxnum}]{eb}. In
order to measure this, we need a sufficient spectral coverage.
Unfortunately, this is not always available because of a lack of 
coverage at X-ray energies for a large number of (low-peaked) faint
LAT sources. For high-peaked blazars we often lack information in the
TeV energy range. 
X-ray data have been taken from the public \Swift archives.
The \textsl{Neil Gehrels Swift Observatory} observes in the 0.3 --
10\,keV energy range. We obtained observations in the time
range of the four-year IceCube period (May 1, 2010 - April 30, 2014),
unless data were only available outside this time range, in which case
these data were used.
For sources without \Swift/XRT observations we used the recently
updated Second \ROSAT All-Sky Survey (2RXS). The available count rates
where converted into fluxes using the factor $1.08\times
10^{-11}$\,erg\,cm$^{-2}$\,s$^{-1}$, which assumes a photon index of
$\Gamma=1.7$ and an absorption of $3\times 10^{20}$\,cm$^{-2}$
\citep{2RXS}.

Several sources that are in the IceCube neutrino uncertainty fields
are not observed by \Swift and were further not detected by ROSAT. For
these sources we used the flux limit of the RASS catalog as an upper
limit. This contributes a large factor to the uncertainties of our
study.

\Swift/XRT data were prepared and extracted with the standard tools
using HEASoft (v. 6.3.2). After running the xrtpipeline to apply
the newest calibration, the target source was extracted using XSELECT
with a circle of 20 pixels (47.146$^{\prime\prime}$), centered on the
source. The background was extracted using an annulus centered on the
source, with radii of 50 (117.866$^{\prime\prime}$) and 70 pixels
(165.012$^{\prime\prime}$), while ensuring that no sources lie in
the background region.

\subsection{3LAC}
In addition to the X-ray data, the \Fermi/LAT data cover the high
energies. As calculating LAT spectra is computationally expensive, we
used the spectra from the 3FGL catalog that are identical in time
integration to the 3LAC catalog \citep{FGL3}. This time range overlaps
by 2.5 years with the IceCube period. Variability in the sources adds
another systematic uncertainty to our study, but we expect that the
effects of more quiescence and flaring combined in all sources will
more or less cancel out, and be small in comparison to modeling
uncertainties. Furthermore, we expect that the 3FGL spectra are a
reasonable approximation of four-year averaged spectra in the IceCube
time period. We note that some (especially weak) sources (e.g.,
1RXS\,J194422.6$-$452326) show a rising component with detections
above 10\,GeV. Such a component can often be explained with background
from the Sun, for instance \citep{FGL3}.

\subsection{Broadband spectral energy distributions }
As a first step, we rebinned the \Swift X-ray data where available. We
rebinned to constrain the spectrum to ten bins, while ensuring that each
bin is above a signal-to-noise ratio of 4.47 to ensure the validity of
$\chi^2$ statistics. In sources where we have a slightly lower
signal-to-noise ratio, we rebinned to a signal-to-noise ratio of 1.5 and used Cash
statistics. Data with fewer photons (rebinning to a signal-to-noise ratio of
1.5 and fewer than 2 bins) were excluded and are not shown in this paper.

In the case of high-peaked sources, the X-ray data sometimes clearly
lie on the synchrotron peak.  Therefore the X-ray data were modeled
with an absorbed power law (tbnew\footnote{available at:
  \url{http://pulsar.sternwarte.uni-erlangen.de/wilms/research/tbabs/}}),
using the fixed Galactic $N_\mathrm{H}$ \citep{Kalberla2005}, the
vern cross-sections \citep{Verner1996} and the wilm abundances
\citep{Wilms2000}. From the best fit to the \Swift/XRT data, we selected
only those spectra with $\Gamma < 2.3$, ensuring that only X-ray data
on the high-energy peak are included in the final SED model. For
ROSAT, only the flux in the full energy band is available, so data
cannot be excluded based on indices.

We constructed the SEDs and then fit them with an absorbed logarithmic
parabola in the \textsl{Interactive Spectral Interpretation System}
\citep[ISIS;][]{Houck2000}. The data were fit in detector space, with the
exception of \textsl{Fermi}/LAT data. Upper limits were not taken
into account for spectral modeling, so in those cases, the best fit is
based solely on \Fermi/LAT data. In a few rare cases, the \Fermi/LAT
spectrum is not consistent with the X-ray index and was not used for
spectral modeling (see, e.g., 1RXS\, J1958.6$-$301119).
The details of the SED modeling approach are described in detail by
\citet{Krauss2016}.

\subsection{Maximum number of neutrinos}
\label{sect-maxnum}
From the best fit to the high-energy SED, we calculated the integrated
energy flux in the energy range 0.1\,keV -- 1\,TeV to ensure that the
range covers the peak of the high-energy component.
We then followed our previous method of calculating the maximum neutrino
number from the integrated energy flux of the high-energy peak of the
broadband spectrum \citep[see][]{eb,Kadler2016}.
We estimated the neutrino energy flux from the electromagnetic energy
flux assuming pion photoproduction and isospin symmetry.
This yields
\begin{equation}
  \begin{aligned}
    \Phi_\nu \stackrel{\mathrm{!}}=&\int\limits_{E_{1,\nu}=30\,\mathrm{TeV}}^{E_{2,\nu}=10\,\mathrm{PeV}} F_{\nu}\,(E_\nu)
        ~\mathrm{d} E_\nu\\
\end{aligned}
,\end{equation}
with the neutrino flux $F_\nu (E_\nu)$ and the integrated
neutrino energy flux $\Phi_\nu$. The energy boundaries of 30\,TeV and
10\,PeV have been based on FSRQs, whose neutrinos spectra are expected
to peak at PeV energies \citep{Mannheim1993}. If the true spectrum
extends lower or higher, the neutrino estimates would be lower.
However, our goal is to estimate whether there is enough power in this
energy band to explain the IceCube neutrinos.
The integrated electromagnetic energy flux is given by 
\begin{equation}
\Phi_\gamma  \stackrel{\mathrm{!}}= \int\limits_{E_{1,\gamma}=0.1\,\mathrm{keV}}^{E_{2,\gamma}=1\,\mathrm{TeV}}
F_{\gamma}\,(E)\,\mathrm{d}E \qquad .
\end{equation}

Numerical evaluation of pion production \citet{Muecke2000} shows that
\begin{equation}
  \Phi_\nu = \Phi_\gamma\qquad,
\label{eq1}
\end{equation}
is a good approximation for most astrophysical
scenarios.
This approach yields a strict upper limit on the flux (or a
maximum possible number of neutrinos that can be detected),
because it assumes that the high-energy peak is purely hadronic and
that the neutrino flux is emitted as a $\delta$-function at the
observed neutrino energy. In the following section, we discuss
corrections to this maximum assumption for the neutrino flux needed to
derive realistic expectation values.

\subsection{Neutrino spectrum correction}
Of course, monoenergetic neutrino production is unrealistic, and a
power-law distribution is a more realistic assumption.
In the proton blazar model, most of the flux is emitted at energies of
1000\,PeV (for BL Lacs) and at $\sim$PeV energies \citep[for
  FSRQs;][]{Biermann1987}; a $\Gamma_\nu=2.58$ spectrum is
inconsistent with both. A pp-component from protons leaking from the
jet and diffusing through the host galaxy is steeper
($\Gamma_\nu=2.3$) but not boosted, and thus comes from the larger
number of radio galaxies. We assume that the integrated
electromagnetic energy flux $\Phi_\gamma$ equals the integrated
neutrino energy flux $\Phi_\nu$, where $E_{1,\gamma}=0.1$\,keV and
$E_{2,\gamma}=1$\,TeV gives the lower and higher energies,
respectively, which were used to obtain the integrated electromagnetic
flux. $E_{1,\nu}=30\,\mathrm{TeV}$ and $E_{2,\nu}=10\,\mathrm{PeV}$
then specify the energy range in which the neutrino spectrum was
defined. The HESE best-fit photon index of $\Gamma_\nu=2.58$ was
obtained from the energy range 60\,TeV--3\,PeV \citep{IC4yr}, although
noting that a spectrum this steep is 
inconsistent with predictions from photo-hadronic source models
\citep{Mannheim1995}.
This can  be used to correct for a more realistic neutrino spectrum.
We assume that the neutrinos follow a power-law flux distribution
\begin{equation}
  F_\nu\, (E_\nu) =  C \cdot E^{1-\Gamma_\nu}_\nu,
  \label{eq2}
\end{equation}
with a photon index of the neutrino spectrum $\Gamma_\nu=2.58$ and a
normalization constant C.
\begin{figure}
  \includegraphics[width=\columnwidth]{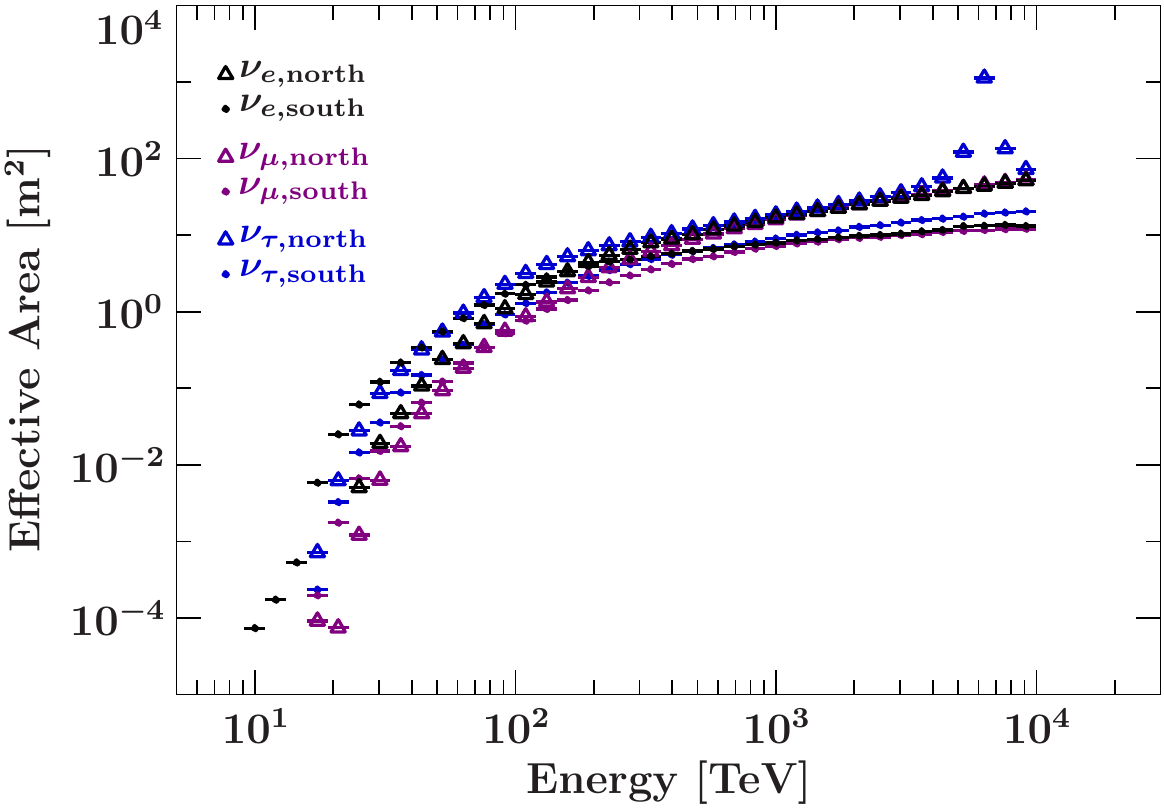}
  \caption{IceCube effective area for all neutrino flavors for the
    northern and southern hemisphere as given by \citet{Icecube2013b}.}
  \label{fig-ar}
\end{figure}
Integrating Eq.~\eqref{eq2} and solving Eq.~\eqref{eq1} for the
normalization constant yields
\begin{equation}
  C = \Phi_\gamma\cdot\dfrac{2 - \Gamma_\nu}{E_{2,\nu}^{2-\Gamma_\nu} -  E_{1,\nu}^{2-\Gamma_\nu} }\qquad.
  \label{eq3}
\end{equation}

The number of observed neutrinos per given energy is given by
\begin{equation}
 \dfrac{\mathrm{d}N_\nu\, (E_\nu)}{\mathrm{d}E_\nu} = \dfrac{F_\nu\,(E_\nu)}{E_\nu} \cdot A_\mathrm{eff}\,(E_\nu) \cdot  T\qquad,
  \label{eq4}
\end{equation}
where $A_\mathrm{eff}$ is the effective area of the neutrino
instrument at a given neutrino energy for either the southern or the
northern hemisphere (see Fig.~\ref{fig-ar}), $T=1347\,$days is the IceCube time range, and
$F_\nu$ is the neutrino flux, dependent on the energy.
Combining Eqs.~\eqref{eq2},~\eqref{eq3}, and~\eqref{eq4} gives
\begin{equation}
\dfrac{\mathrm{d}N_{\nu}\,(\Gamma_\nu, E_\nu)}{\mathrm{d} E_\nu} =
\dfrac{\Phi_\gamma\cdot(2-\Gamma_\nu)}{E_{2,\nu}^{2-\Gamma_\nu}-E_{1,\nu}^{2-\Gamma_\nu}}
\cdot E^{-\Gamma_\nu}_\nu\cdot T \cdot A_\mathrm{eff}\,(E_\nu)\qquad,
\label{eq-finals}
\end{equation}
where $E_\nu$ is the energy of the detected neutrino event.
In order to obtain the total number of neutrinos per source in the
energy range 100\,TeV--1\,PeV, we integrated Eq.~\eqref{eq-finals}.
As the IceCube effective area is dependent on energy (see
Fig.~\ref{fig-ar}), we integrated numerically over each energy bin of the effective
area, where $A_\mathrm{eff}\,(E_\nu)$ is constant. The integral of one single
bin at the neutrino energy $E^\prime_\nu$ is given by

\begin{equation}
  \begin{aligned}
    N_\nu = \int\limits_{E_1^\prime}^{E_2^\prime} & \dfrac{dN_\nu\, (E_\nu)}{dE_\nu}~\mathrm d E_\nu\\=
& \left.\dfrac{C\cdot T}{1-\Gamma_\nu}\cdot A_\mathrm{eff}\,(E_\nu^\prime)\cdot E_\nu^{1-\Gamma_\nu} ~
\right|_{E_{1^\prime}}^{E_{2^\prime}}\\
= &
\dfrac{\Phi_\gamma\cdot(2-\Gamma_\nu)}{(E_{2,\nu}^{2-\Gamma_\nu}-E_{1,\nu}^{2-\Gamma_\nu})\cdot
(1-\Gamma_\nu)} \cdot T \cdot A_\mathrm{eff}\,(E^\prime_\nu) \cdot \left(E_{2^\prime}^{1-\Gamma_\nu}-E_{1^\prime}^{1-\Gamma_\nu}\right)
\end{aligned}
.\end{equation}

The resulting neutrino numbers are given in Table \ref{tab-nunum} in the
sixth column ($\nu_\mathrm{scale,int}$), but we note that this does
not give the expected number of neutrinos per neutrino event, but per
source in the whole energy range.
We note that the total number of
scaled, integrated neutrinos is representative for the total expected
number of neutrinos from blazars.
We note that the true correction factors are smaller if a fraction of
the energy flux lies outside of the chosen window (e.g., BL Lac
emission above 10\,PeV, if the proton blazar model is correct, and
below 30\,TeV if a steep spectrum with $\Gamma_\nu=2.58$ is true).
Furthermore, such a steep spectrum cannot be true to below a few TeV
without overproducing $\gamma$ rays, or overproducing the nonthermal
output of jets.

\section {Results}
\begin{figure}
  \includegraphics[width=\columnwidth]{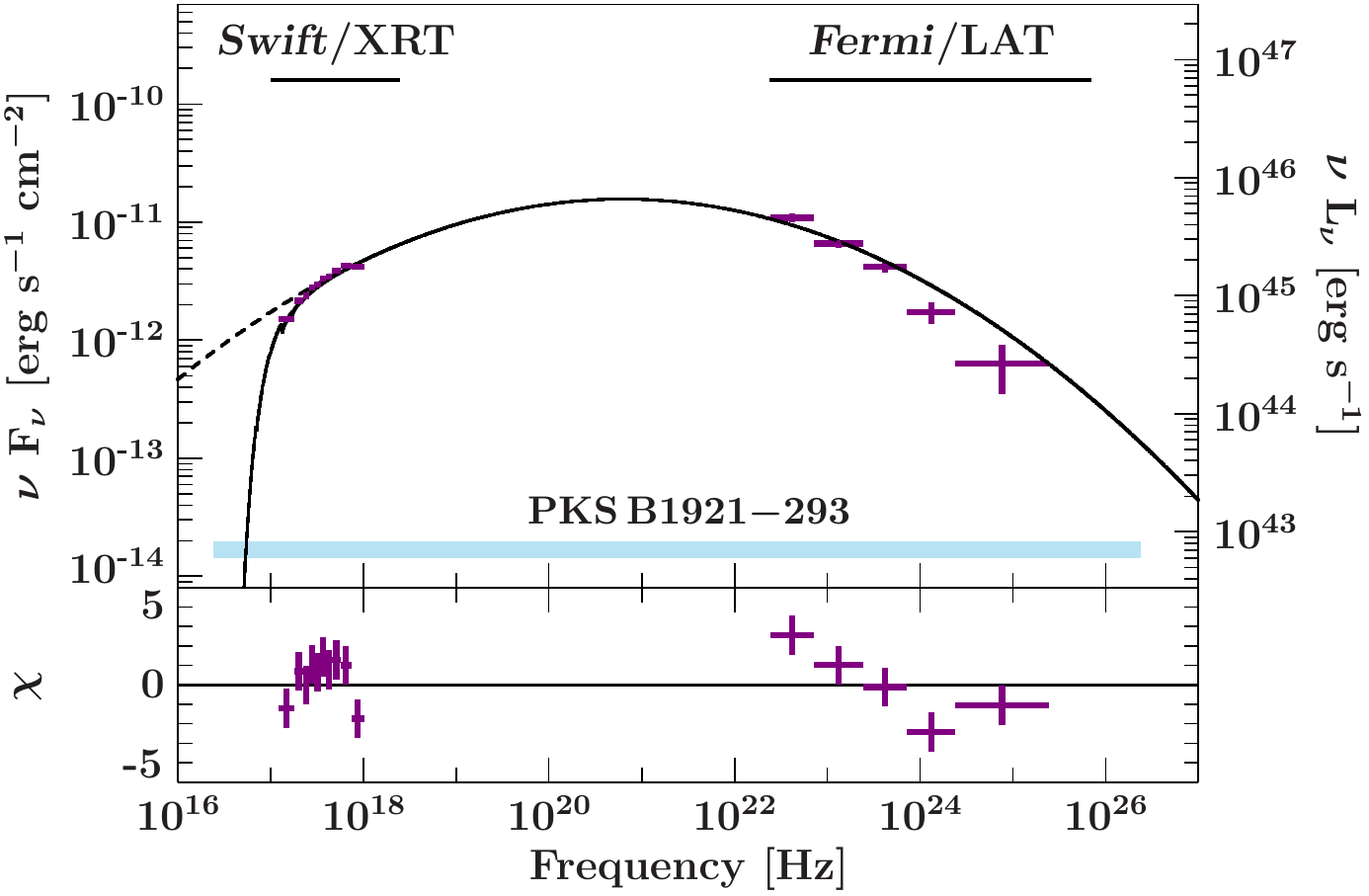}
  \caption{Example SED with good \Swift/XRT data that is well
    described by a logarithmic parabola (solid line). The unabsorbed
    model is shown as a dashed line. Residuals are shown in the lower
    panel. The energy range that is used for integrating the
    electromagnetic flux is given by the blue bar. The SEDs of the
    other sources are given in Fig.~\ref{fig-sed0}--Fig.~\ref{fig-sed15}}
  \label{fig-sedn}
\end{figure}

There are 179 3LAC sources consistent with IceCube HESE events between
100\,TeV and 1\,PeV.
We find that most of the sources can be described well by the model
(see Fig.~\ref{fig-sedn}),
and when this is not true, it is mostly due to a lack of
data. In Fig.~\ref{fig-sed0}--Fig.\ref{fig-sed15} we show the SEDs of
all sources including their best-fit model. Here we discuss the
results obtained from the neutrino calculation.

\begin{figure}
\resizebox{\hsize}{!}{\includegraphics[width=0.46\textwidth]{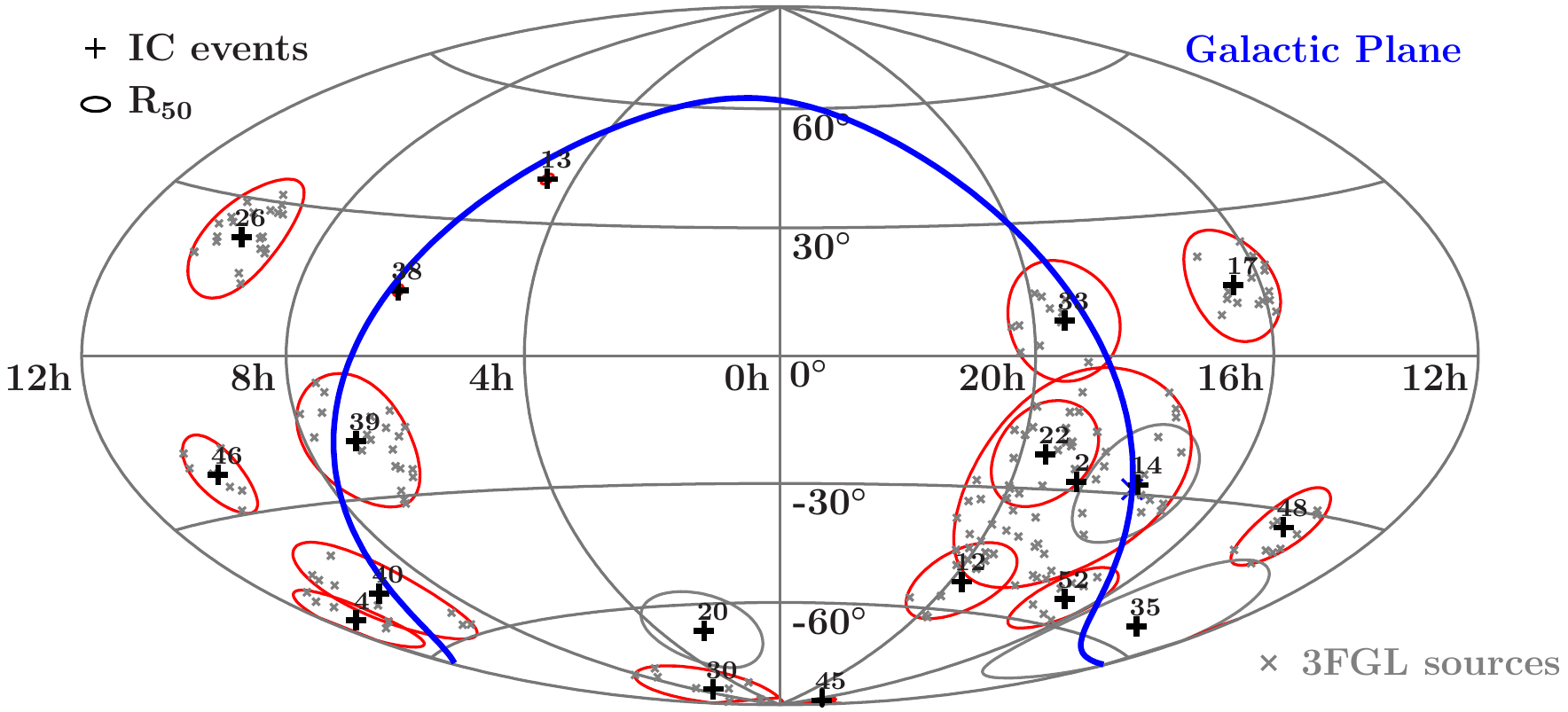}}
\caption{Hammer-Aitoff projection of the sky in right ascension and
  declination including all HESE events above 100\,TeV (red), and above
  1\,PeV (gray) in four years of data, with their respective $R_{50}$
  error circles. The 3FGL position of the 3LAC sources consistent with
the error circles are shown in gray crosses and the number of the
IceCube event is marked above the plus sign, which marks the best-fit
position of the origin of the neutrino event.}
\label{fig-reg}
\end{figure}

The resulting neutrino numbers are given in Table~\ref{tab-nunum},
where the sources that are the strongest neutrino producers are listed
first\footnote{We note that the expected number of neutrinos
  for the source SUMSS\,J074220$-$813139 (IC\,30) has been manually
  set to zero, as this source has no detections in either the
  \Fermi/LAT energy bands or in the X-rays, and we therefore consider
  any estimate of the flux to be highly unreliable.}.
We note that in some fields the predicted number of neutrinos is
dominated by the brightest sources, while in others the expected
output is distributed fairly evenly across the individual sources.

The maximum number of neutrinos obtained for all sources is
$N_{\mathrm{max,all}} = \totaln$, which
clearly exceeds the 16 detected neutrinos between 100\,TeV and 1\,PeV.
Correcting this for the neutrino spectrum and integrating lowers this
value, yielding a total of \intn~neutrinos. This is still larger than the
detected number of neutrinos. Applying the empirical factor from
\citet{Kadler2016} yields a total of 4 neutrinos.

Given the contributions from atmospheric neutrinos in this sample,
this suggests that blazars could contribute largely to the observed
neutrino signal in this energy range.
However, our calculations show that the energy range over which
blazars emit neutrinos and the spectral shape strongly affect neutrino
estimates.

To calculate the number of neutrinos for
individual events, we used all sources that are consistent with the uncertainty
region of the neutrino event. To calculate the total number,
we exclude sources that appeared several times, as the uncertainty
regions overlap (see Fig.~\ref{fig-reg}).

\subsection{Scaling factor}

Another aspect to consider is which properties of the
sources can further lower the neutrino number (in addition to the neutrino
spectrum).
\citet{Kadler2016} used an empirical scaling factor, comparing the
observed event to the expected maximum number of neutrinos. This
yields an empirical scaling of $\sim 0.0089$.
Based on \citet{IC4yr}, we assumed 10\% atmospheric neutrino events above
100\,TeV, that is to say, two cosmic events. The empirical scaling factor is then
given by $f_\mathrm{emp}=14/\totaln=0.028$, which is close to
$f_\mathrm{emp}=0.0089$ that we derived from the IC35 field alone
\citep{Kadler2016}.

The empirical scaling factor is explained by a combination of neutrino
flavor, blazar class, and spectrum factor. We expect to not detect the
neutrino flavors in IceCube data equally well
\citep{Learned1995,Aartsen2015}, which would further reduce the number
of detected neutrinos by $\sim1/2$, which, even with the spectrum
correction, would be in excess of the detected neutrino events.

It is unclear if all blazars contribute equally, that is, if FSRQs are
more or less hadronic than BL Lac-type objects. This could account for
the second factor ($\sim 1/2$). The third factor is the effect of the neutrino
spectrum. As we already took this into account, we added another factor.
Assuming a more realistic lepto-hadronic SED than in our
previous approach, that is, a high-energy peak that is not purely
hadronic, we further lowered the number of neutrinos. As we do not know
how hadronic blazars are, the exact value remains uncertain.
Here we assumed that all 14 cosmic neutrinos in the energy range were
produced by blazars (including FSRQs and BL Lacs). Applying the
neutrino flavor factor and the spectral correction yields an expected
number of $\intn\cdot0.5\cdot0.5=\intnhalfhalf$~neutrino events. To
lower this to the detected number of events requires only \perchadr\%
of the high-energy emission to originate from hadronic interactions.
This estimate neglects contributions from other sources and from unresolved blazars.

\subsection{Full-sky approximation}
While we only calculated the SED for the sources that areconsistent with the
neutrino events, in search for possible excesses, the full sky
contribution must be considered.
This effect of overestimating the neutrino flux is related to the
Eddington bias \citep{Strotjohann2018}, which overestimates the neutrino
flux when only  a small number of events is considered.
The neutrino expectations were
calculated for all 16 neutrio uncertainty regions, which cover an area
of 5590.88\,degrees$^2$. If we assume that the neutrino uncertainty
regions are representative of the rest of the sky, we can scale the
calculated neutrino numbers to the full sky (41253\,degrees$^2$). The
resulting total number of neutrinos is $N_{\mathrm{max,all,fullsky}} =
\totalnf$, which yields a scaled number of neutrinos
$N_{\mathrm{scale,fullsky}}=\srcscalf$. Applying the other factors
from \citet{Kadler2016} yields $N_\mathrm{f, fullsky}=\totalf$.
Comparing these numbers to the observed 14 cosmic neutrinos yields that
only \perchadrf\% of blazar high-energy emission is of hadronic
origin, for blazars to explain all cosmic neutrinos in the
100\,TeV--10\,PeV range. If this is true, a correlation between
high-energy electromagnetic and neutrino flux cannot be expected.

\subsection{Effect of the neutrino spectrum}
We assumed a power law for the neutrino spectrum, which might be an incorrect
assumption. Furthermore, here we assumed that all neutrinos from a source
are produced within the given energy range of 100\,TeV--10\,PeV,
which neglects higher- and lower-energy neutrinos.
The proton blazar model predicts a hard spectrum \citep{Mannheim1995},
which is inconsistent with the measured IceCube index of $\Gamma=2.58$.
Even when this soft spectrum is used, calorimetrically blazars can explain
the observed neutrino events above 100\,TeV.

\subsection{Energy range}

\begin{figure}
  \includegraphics[width=\columnwidth]{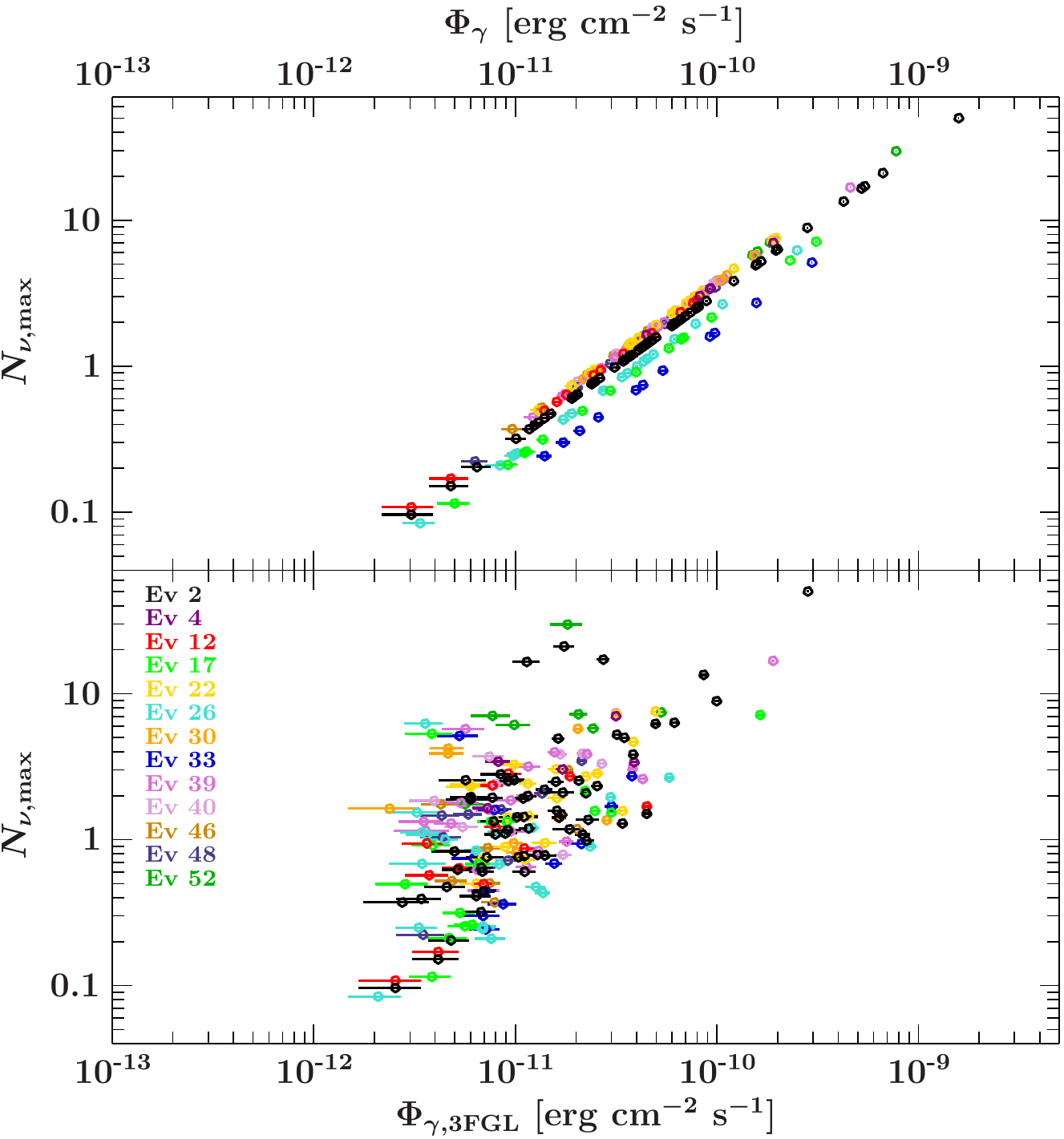}
  \caption{Calculated maximum number of neutrinos vs.
  the total integrated model flux between 1\,keV and 1\,PeV (top) and
  the calculated number of neutrinos vs. the given 3FGL flux. This
  illustrates that the \Fermi/LAT flux alone is not a good predictor
  of expected neutrino numbers.}
  \label{fig-fglnu}
\end{figure}
The number of neutrinos that might be produced by the sources listed
here depends on many factors, such as the selected neutrino spectrum.
However, the basic input is the integrated electromagnetic flux of the
source. The specific energy range of integration does not affect
the expected number of neutrinos (as long as the peak is covered).
This shows, however, that the expected number of neutrinos is
dependent on the total integrated high-energy flux (assuming it is
purely hadronic), not on the \Fermi/LAT flux alone, for instance.
This is illustrated in Fig.~\ref{fig-fglnu}, where we compare the 3FGL
energy flux with the expected number of neutrinos.

\section{Discussion}
\label{sec-caveat}
In this section we discuss possible problems with our method and
other caveats. We discuss the high number of neutrinos that were
calculated, as well as the impact of blazar variability on the
neutrino number. We also discuss the problems related to the sparse
broadband coverage and faint sources.

\subsection{High neutrino numbers}
The highest maximum number of neutrinos has been predicted for the source
PKS\,1830$-$211, with \srctot~predicted neutrinos. This is due to its
steep spectra and the subsequent parabola, which peaks at high fluxes in
the MeV band. However, when we scale this value and others to a
realistic neutrino spectrum, the predicted neutrino number falls to
\srcscal~expected events for this source in the 100\,TeV--1\,PeV energy
range, and to 3 neutrinos when the other scaling factors are applied.
This is still above the detected number of neutrinos. However, when
we take into account that most sources are expected to have a strong
or even dominant leptonic contribution, this values becomes much lower,
consistent with a non-detection. It is further unclear whether some
sources are purely leptonic, while others have a strong hadronic
component, or if this ``hadronicness'' is similar in all blazars.

\subsection{Blazar variability}
For many sources in our sample, multiwavelength observations are
sparse, especially for fainter and less well-known sources, and
simultaneous data are often not available. In such cases, the
individual narrow-band observations from different epochs combined
into an SED might not be representative of the true average SED
because of source variability between the epochs.
As many of the sources are faint and barely detected by \Fermi/LAT, it
is impossible to generate light curves for many of these sources in
order to study their variability. Furthermore, multiple observations or
monitoring in other wavebands is only available for a very small
number of sources.

\subsection{Broadband coverage}
As observations are sparse, the SED coverage is far from ideal. Many of the SEDs are constrained by only \ROSAT upper limits, as
no \Swift/XRT pointings are available or the signal-to-noise ratio is not
sufficient for spectral fitting. Furthermore, several \Fermi-LAT
spectra have upper limits, yielding only one or two energy bands with
detections (TS\,$>25$) in extreme cases.
This sparse coverage, as well as neglecting the synchrotron peak,
leads to over- or underestimates of the electromagnetic flux and therefore
of the number of neutrinos. We estimate that this could change the
integrated flux to a factors of a few, but would not greatly affect the total
number of neutrinos.

\subsection{Faint sources}
As neutrinos have a very long mean free path, we can observe neutrinos
from objects that are too faint to be resolved with current
instruments. We therefore estimate additional neutrinos from faint, unresolved
blazars. A rough estimate can be obtained from the extragalactic
$\gamma$-ray background (EGB). Between 50\% and 80\% of the EGB is
expected to be from blazars \citep{egb1,egb2}. This total integrated
flux from unresolved blazars is a factor of 2 higher than the total
integrated value for resolved blazars \citep{Krauss2015}. Furthermore,
unbeamed sources, such as radio galaxies, are also expected to
contribute to the neutrino (and to the extragalactic $\gamma$-ray
background).

\subsection{Source classes}
Blazars can further be separated into \mbox{FSRQs} and BL Lacs depending on
their jet power.
While \mbox{FSRQs} are the more likely counterparts for neutrinos in
the IceCube energy range \citep[due to interactions with thermal UV
  photons from the accretion disk][]{Mannheim1993b}, some authors have considered only 
BL Lac objects \citep{Righi2017}. It is unclear how the
opacity problem in these sources can be overcome since BL Lacs are generally optically
thin to the absorption of TeV $\gamma$ rays and thus are very inefficient in
producing neutrinos at IceCube energies \citep{Bahcall2001}.
However, possible exceptions are being discussed, for example by \citet{Ansoldi2018, Murase2018}.

\section{Conclusions}
We have constructed SEDs for all 3LAC counterparts to IceCube HESE
neutrino events between 100\,TeV and 1\,PeV. We also showed that the
\Fermi/LAT energy flux does not yield a good estimate of the neutrino
flux.
From the integrated energy flux, we estimated the maximum number of
neutrino events from all blazars $N_\nu = \totaln$, which is in
agreement with our previous work, where we found that blazars are easily able to
explain the neutrino signal based on their electromagnetic output at
high energies \citep{eb}.
When a power-law neutrino spectrum with a spectral index
$\Gamma_\nu=2.58$ is adopted, this number is reduced to $\intn$~neutrinos events.
These numbers also show no strong neutrino excesses from individual
sources ($>3$ neutrino events) when the other scaling
factors are applied, consistent with the non-detection of high-energy neutrino
multiplets from bright sources.

Applying the empirical factor from \citet{Kadler2016} to the maximum
number of neutrino events instead of using the correction to the
neutrino spectrum reduces this to four neutrino events.
Another factor that could reduce the number of neutrinos would be the
``hadronicness'' of the source, which accounts for the fraction of the
high-energy emission that is actually due to photo-meson production.
We find that if \perchadr\% of the high-energy electromagnetic flux
originates from hadronic emission, this is sufficient to explain the
IceCube neutrino events above $100\,$TeV.
When these numbers are scaled to the full sky, we obtain a total number of
neutrinos $N_{\mathrm{max,all,fullsky}} = \totalnf$, the scaled number
of neutrinos $N_{\mathrm{scale,fullsky}}=\srcscalf$, and $N_\mathrm{f,
  fullsky}=\totalf$. Comparing these numbers to the observed 14 cosmic
neutrinos yields that only \perchadrf\% of blazar high-energy emission
is of hadronic origin, for blazars to explain all cosmic neutrinos in
the 100\,TeV--10\,PeV range. If this is true, a correlation between
high-energy electromagnetic and neutrino flux cannot be expected.

Considering the systematic uncertainties as well as the caveats, we
argue that blazars can contribute strongly to the 100\,TeV--1\,PeV
energy range. Considering a realistic neutrino spectrum and other
aspects of blazar physics reduces the amount of neutrinos we expect to
detect from these sources, but is still larger than the detected
number of neutrinos and constrains the ``hadronicness'' of blazars. 
The recently detected TXS\,0506+056 (which is not a HESE event and not
detected in four years of data), which is in agreement with the track
event IC\,170922A, is a tantalizing hint at a blazar connection, which
is why further study is necessary \citep{txs}.

\begin{acknowledgements}
We thank A. Heijboer for the useful discussions regarding the results.
F. K. acknowledges funding from the European Union’s Horizon 2020
research and innovation programme under grant agreement No 653477.
B. W. acknowledges funding from the DAAD RISE program.

This research has made use of ISIS functions (ISISscripts) provided by 
ECAP/Remeis observatory and MIT
(\url{http://www.sternwarte.uni-erlangen.de/isis/}).
\end{acknowledgements}

\bibliographystyle{jwaabib}
\bibliography{mnemonic,aa_abbrv,all}
\renewcommand*{\arraystretch}{1}

\clearpage
\newpage
\begin{appendix}
  \section{Number of neutrinos}
\begin{table}[b]\centering\fontsize{5.5}{5.7}\selectfont
\caption{Neutrino numbers for all sources. This table lists the IceCube HESE event number in the first column, the maximum number of neutrinos in the second column, the neutrino number scaled to the empirical factor in the third column, and the neutrino numbers corrected for the realistic neutrino spectrum in the fourth column, integrated over the full energy range (which does not take into account that other neutrinos may have been seen from a given source).The last column shows the values with the blazar and neutrino flavor scalings applied. Within each neutrino event, sources are ordered by their predicted number of neutrinos. Neutrinos from sources that appear multiple times because of overlap of their uncertainty regions are marked with parentheses the second time, and only the first of their neutrino expectations are counted.}
\label{tab-nunum}
\begin{tabular}{llllll}

IC & Source & \#$\nu$ & \#$\nu_{\mathrm{emp.}}$ & \#$\nu_{\mathrm{scale, int.}}$ & \#$\nu_{\mathrm{f}}$\\
\hline
2 & PKS1830-211 & 50.18 & 4.47e-01 & 1.07e+01 & 2.68e+00\\
 & 1RXSJ174459.5-172640 & 21.10 & 1.88e-01 & 4.51e+00 & 1.13e+00\\
 & SwiftJ1656.3-3302 & 17.14 & 1.53e-01 & 3.66e+00 & 9.15e-01\\
 & 1RXSJ182853.8-241746 & 16.52 & 1.47e-01 & 3.53e+00 & 8.82e-01\\
 & TXS1714-336 & 13.48 & 1.20e-01 & 2.88e+00 & 7.19e-01\\
 & PMNJ1802-3940 & 8.90 & 7.92e-02 & 1.90e+00 & 4.75e-01\\
 & PKS1730-13 & 6.33 & 5.64e-02 & 1.35e+00 & 3.38e-01\\
 & PKSB1908-201 & 6.22 & 5.53e-02 & 1.33e+00 & 3.32e-01\\
 & PKSB1921-293 & 5.24 & 4.66e-02 & 1.12e+00 & 2.80e-01\\
 & PKS1954-388 & 4.99 & 4.45e-02 & 1.07e+00 & 2.67e-01\\
 & PKS1933-400 & 4.93 & 4.39e-02 & 1.05e+00 & 2.63e-01\\
 & TXS1920-211 & 3.84 & 3.42e-02 & 8.20e-01 & 2.05e-01\\
 & PMNJ1912-0804 & 2.80 & 2.50e-02 & 5.99e-01 & 1.50e-01\\
 & PMNJ2000-2931 & 2.56 & 2.28e-02 & 5.47e-01 & 1.37e-01\\
 & 1RXSJ171405.2-202747 & 2.56 & 2.28e-02 & 5.46e-01 & 1.37e-01\\
 & PMNJ1959-4246 & 2.54 & 2.26e-02 & 5.42e-01 & 1.36e-01\\
 & PMNJ1936-4719 & 2.52 & 2.24e-02 & 5.38e-01 & 1.35e-01\\
 & PMNJ1858-2511 & 2.49 & 2.22e-02 & 5.33e-01 & 1.33e-01\\
 & TXS1951-115 & 2.33 & 2.07e-02 & 4.98e-01 & 1.24e-01\\
 & PMNJ1758-4820 & 2.20 & 1.96e-02 & 4.70e-01 & 1.17e-01\\
 & PMNJ1718-3056 & 2.10 & 1.87e-02 & 4.49e-01 & 1.12e-01\\
 & PKS1958-179 & 2.08 & 1.85e-02 & 4.44e-01 & 1.11e-01\\
 & TXS2002-233 & 1.99 & 1.77e-02 & 4.24e-01 & 1.06e-01\\
 & 1RXSJ195815.6-301119 & 1.95 & 1.74e-02 & 4.17e-01 & 1.04e-01\\
 & PKS1929-457 & 1.94 & 1.73e-02 & 4.14e-01 & 1.04e-01\\
 & PKS2021-330 & 1.92 & 1.71e-02 & 4.09e-01 & 1.02e-01\\
 & 1RXSJ184919.7-164726 & 1.89 & 1.68e-02 & 4.04e-01 & 1.01e-01\\
 & TXS1909-124 & 1.58 & 1.41e-02 & 3.37e-01 & 8.43e-02\\
 & PKS2005-489 & 1.50 & 1.34e-02 & 3.21e-01 & 8.03e-02\\
 & PMNJ1830-4441 & 1.49 & 1.32e-02 & 3.18e-01 & 7.94e-02\\
 & NVSSJ165949-313047 & 1.43 & 1.27e-02 & 3.06e-01 & 7.64e-02\\
 & PMNJ1735-1117 & 1.43 & 1.27e-02 & 3.04e-01 & 7.61e-02\\
 & PKS1821-525 & 1.41 & 1.26e-02 & 3.02e-01 & 7.55e-02\\
 & NVSSJ173146-300309 & 1.37 & 1.22e-02 & 2.92e-01 & 7.30e-02\\
 & 1RXSJ194306.8-351001 & 1.33 & 1.19e-02 & 2.84e-01 & 7.11e-02\\
 & 1H1914-194 & 1.29 & 1.15e-02 & 2.76e-01 & 6.89e-02\\
 & PKS1942-313 & 1.19 & 1.06e-02 & 2.55e-01 & 6.37e-02\\
 & NVSSJ174154-253743 & 1.18 & 1.05e-02 & 2.52e-01 & 6.29e-02\\
 & PMNJ1911-1908 & 1.16 & 1.04e-02 & 2.48e-01 & 6.21e-02\\
 & 1RXSJ195500.6-160328 & 1.10 & 9.80e-03 & 2.35e-01 & 5.88e-02\\
 & PKS2004-447 & 1.09 & 9.67e-03 & 2.32e-01 & 5.80e-02\\
 & PMNJ1918-4111 & 1.08 & 9.63e-03 & 2.31e-01 & 5.78e-02\\
 & NVSSJ182338-345412 & 0.98 & 8.75e-03 & 2.10e-01 & 5.25e-02\\
 & PMNJ1808-5011 & 0.83 & 7.39e-03 & 1.77e-01 & 4.43e-02\\
 & PMNJ1849-4314 & 0.79 & 7.00e-03 & 1.68e-01 & 4.20e-02\\
 & PMNJ1921-1607 & 0.78 & 6.94e-03 & 1.67e-01 & 4.16e-02\\
 & SUMSSJ195945-472519 & 0.77 & 6.89e-03 & 1.65e-01 & 4.13e-02\\
 & PMNJ1753-5015 & 0.76 & 6.75e-03 & 1.62e-01 & 4.05e-02\\
 & PMNJ1831-2714 & 0.76 & 6.74e-03 & 1.62e-01 & 4.04e-02\\
 & PMNJ2036-2830 & 0.64 & 5.73e-03 & 1.37e-01 & 3.43e-02\\
 & PMNJ1816-4943 & 0.62 & 5.54e-03 & 1.33e-01 & 3.32e-02\\
 & PMNJ2012-1646 & 0.60 & 5.38e-03 & 1.29e-01 & 3.22e-02\\
 & TXS1918-126 & 0.60 & 5.37e-03 & 1.29e-01 & 3.22e-02\\
 & NVSSJ204150-373341 & 0.47 & 4.22e-03 & 1.01e-01 & 2.53e-02\\
 & PMNJ2022-4513 & 0.44 & 3.94e-03 & 9.45e-02 & 2.36e-02\\
 & PKS1953-325 & 0.41 & 3.66e-03 & 8.77e-02 & 2.19e-02\\
 & 1RXSJ201731.2-411452 & 0.39 & 3.50e-03 & 8.40e-02 & 2.10e-02\\
 & 1RXSJ203650.9-332817 & 0.37 & 3.31e-03 & 7.93e-02 & 1.98e-02\\
 & PMNJ1913-3630 & 0.32 & 2.84e-03 & 6.82e-02 & 1.71e-02\\
 & PMNJ1848-4230 & 0.20 & 1.82e-03 & 4.36e-02 & 1.09e-02\\
 & SUMSSJ193946-492539 & 0.15 & 1.35e-03 & 3.23e-02 & 8.09e-03\\
 & 1RXSJ194422.6-452326 & 0.10 & 8.58e-04 & 2.06e-02 & 5.15e-03\\
\hline
& & \textbf{223.40} & \textbf{2.0e+00} &\textbf{47.7}  & \textbf{1.2e+01} \\
\hline
\end{tabular}
\end{table}
\begin{table}[b]\centering\fontsize{5.5}{5.7}\selectfont
\addtocounter{table}{-1}
\caption{contd.}
\begin{tabular}{llllll}

IC & Source & \#$\nu$ & \#$\nu_{\mathrm{emp.}}$ & \#$\nu_{\mathrm{scale, int.}}$ & \#$\nu_{\mathrm{f}}$\\
\hline
4 & MRC1036-529 & 7.02 & 6.24e-02 & 1.29e+00 & 3.22e-01\\
 & PKS1116-46 & 3.42 & 3.05e-02 & 6.29e-01 & 1.57e-01\\
 & PKS1101-536 & 3.40 & 3.02e-02 & 6.24e-01 & 1.56e-01\\
 & PKS1104-445 & 3.03 & 2.70e-02 & 5.57e-01 & 1.39e-01\\
 & PMNJ1109-4815 & 1.64 & 1.46e-02 & 3.02e-01 & 7.56e-02\\
\hline
& & \textbf{18.51} & \textbf{1.6e-01} &\textbf{3.4}  & \textbf{8.5e-01} \\
\hline
12 & PMNJ1936-4719 & 2.83 & 2.52e-02 & 5.38e-01 & 1.35e-01\\
 & (PKS1936-623) & 2.71 & 2.42e-02 & 5.16e-01 & 1.29e-01\\
 & PKS1929-457 & 2.35 & 2.09e-02 & 4.46e-01 & 1.12e-01\\
 & PKS2005-489 & 1.69 & 1.51e-02 & 3.21e-01 & 8.03e-02\\
 & (PKS2004-447) & 1.22 & 1.09e-02 & 2.32e-01 & 5.80e-02\\
 & PMNJ1949-6137 & 0.94 & 8.39e-03 & 1.79e-01 & 4.48e-02\\
 & (SUMSSJ195945-472519) & 0.87 & 7.74e-03 & 1.65e-01 & 4.13e-02\\
 & (1RXSJ195503.1-564031) & 0.64 & 5.67e-03 & 1.21e-01 & 3.03e-02\\
 & (PKS2036-577) & 0.57 & 5.08e-03 & 1.08e-01 & 2.71e-02\\
 & (PMNJ2022-4513) & 0.50 & 4.43e-03 & 9.45e-02 & 2.36e-02\\
 & (SUMSSJ193946-492539) & 0.17 & 1.52e-03 & 3.23e-02 & 8.09e-03\\
 & (1RXSJ194422.6-452326) & 0.11 & 9.64e-04 & 2.06e-02 & 5.15e-03\\
\hline
& & \textbf{14.61} & \textbf{1.3e-01} &\textbf{2.8}  & \textbf{6.9e-01} \\
\hline
13& NONE & 0.0 & 0.0 & 0.0\\
\hline
& & \textbf{0.00} & \textbf{0.0e+00} &\textbf{0.0}  & \textbf{0.0e+00} \\
\hline
17 & PG1553+113 & 7.15 & 6.36e-02 & 1.26e+00 & 3.14e-01\\
 & 1RXSJ163717.1+131418 & 5.31 & 4.73e-02 & 9.33e-01 & 2.33e-01\\
 & 4C+10.45 & 2.16 & 1.93e-02 & 3.80e-01 & 9.50e-02\\
 & 4C+15.54 & 1.57 & 1.40e-02 & 2.76e-01 & 6.91e-02\\
 & PKS1551+130 & 1.53 & 1.36e-02 & 2.69e-01 & 6.73e-02\\
 & MG1J163119+1051 & 1.33 & 1.18e-02 & 2.33e-01 & 5.84e-02\\
 & MG1J165034+0824 & 1.33 & 1.18e-02 & 2.33e-01 & 5.84e-02\\
 & 1ES1552+203 & 0.91 & 8.13e-03 & 1.60e-01 & 4.01e-02\\
 & MG2J165546+2043 & 0.68 & 6.06e-03 & 1.20e-01 & 2.99e-02\\
 & MG1J154930+1708 & 0.50 & 4.41e-03 & 8.70e-02 & 2.17e-02\\
 & TXS1549+089 & 0.31 & 2.80e-03 & 5.53e-02 & 1.38e-02\\
 & MG1J160340+1106 & 0.26 & 2.32e-03 & 4.59e-02 & 1.15e-02\\
 & 87GB155744.0+232525 & 0.26 & 2.27e-03 & 4.48e-02 & 1.12e-02\\
 & TXS1638+118 & 0.21 & 1.88e-03 & 3.71e-02 & 9.28e-03\\
 & MG1J154628+1817 & 0.11 & 1.02e-03 & 2.02e-02 & 5.05e-03\\
\hline
& & \textbf{23.64} & \textbf{2.1e-01} &\textbf{4.2}  & \textbf{1.0e+00} \\
\hline
22 & (PKSB1908-201) & 7.57 & 6.74e-02 & 1.33e+00 & 3.32e-01\\
 & (PKSB1921-293) & 7.21 & 6.42e-02 & 1.26e+00 & 3.16e-01\\
 & (TXS1920-211) & 4.68 & 4.16e-02 & 8.20e-01 & 2.05e-01\\
 & (PMNJ2000-2931) & 3.26 & 2.90e-02 & 5.72e-01 & 1.43e-01\\
 & (PMNJ1858-2511) & 3.04 & 2.71e-02 & 5.33e-01 & 1.33e-01\\
 & (TXS1951-115) & 2.84 & 2.53e-02 & 4.98e-01 & 1.24e-01\\
 & (PKS1958-179) & 2.71 & 2.41e-02 & 4.75e-01 & 1.19e-01\\
 & (TXS2002-233) & 2.42 & 2.15e-02 & 4.24e-01 & 1.06e-01\\
 & (1RXSJ195815.6-301119) & 2.38 & 2.12e-02 & 4.17e-01 & 1.04e-01\\
 & (1RXSJ184919.7-164726) & 2.30 & 2.05e-02 & 4.04e-01 & 1.01e-01\\
 & (TXS1909-124) & 1.92 & 1.71e-02 & 3.37e-01 & 8.43e-02\\
 & (1H1914-194) & 1.57 & 1.40e-02 & 2.76e-01 & 6.89e-02\\
 & (PKS1942-313) & 1.45 & 1.29e-02 & 2.55e-01 & 6.37e-02\\
 & (PMNJ1911-1908) & 1.42 & 1.26e-02 & 2.48e-01 & 6.21e-02\\
 & (PMNJ1921-1607) & 0.95 & 8.46e-03 & 1.67e-01 & 4.16e-02\\
 & (1RXSJ195500.6-160328) & 0.89 & 7.94e-03 & 1.56e-01 & 3.91e-02\\
 & (PMNJ2012-1646) & 0.74 & 6.55e-03 & 1.29e-01 & 3.22e-02\\
 & (TXS1918-126) & 0.73 & 6.52e-03 & 1.28e-01 & 3.21e-02\\
 & (PKS1953-325) & 0.50 & 4.45e-03 & 8.77e-02 & 2.19e-02\\
\hline
& & \textbf{48.59} & \textbf{4.3e-01} &\textbf{8.5}  & \textbf{2.1e+00} \\
\hline
\end{tabular}
\end{table}
\begin{table}[b]\centering\fontsize{5.5}{5.7}\selectfont
\addtocounter{table}{-1}
\caption{contd.}
\begin{tabular}{llllll}

IC & Source & \#$\nu$ & \#$\nu_{\mathrm{emp.}}$ & \#$\nu_{\mathrm{scale, int.}}$ & \#$\nu_{\mathrm{f}}$\\
\hline
26 & RXJ0959.4+2123 & 6.24 & 5.55e-02 & 1.01e+00 & 2.52e-01\\
 & OJ287 & 2.66 & 2.37e-02 & 4.31e-01 & 1.08e-01\\
 & MG2J101241+2439 & 1.96 & 1.74e-02 & 3.17e-01 & 7.92e-02\\
 & 1RXSJ091211.9+275955 & 1.54 & 1.37e-02 & 2.49e-01 & 6.21e-02\\
 & OK290 & 1.21 & 1.07e-02 & 1.95e-01 & 4.89e-02\\
 & TXS0907+230 & 1.13 & 1.01e-02 & 1.83e-01 & 4.57e-02\\
 & 1RXSJ100235.8+221609 & 1.12 & 1.00e-02 & 1.82e-01 & 4.55e-02\\
 & GB6J0905+2748A & 1.08 & 9.63e-03 & 1.75e-01 & 4.38e-02\\
 & MG2J094148+2728 & 1.00 & 8.91e-03 & 1.62e-01 & 4.05e-02\\
 & Ton0396 & 0.90 & 7.99e-03 & 1.45e-01 & 3.63e-02\\
 & B20920+28 & 0.84 & 7.51e-03 & 1.37e-01 & 3.41e-02\\
 & SDSSJ091230.61+155528.0 & 0.68 & 6.07e-03 & 1.10e-01 & 2.76e-02\\
 & GB6J1001+2911 & 0.68 & 6.05e-03 & 1.10e-01 & 2.75e-02\\
 & MG1J090534+1358 & 0.48 & 4.23e-03 & 7.69e-02 & 1.92e-02\\
 & NVSSJ090226+205045 & 0.43 & 3.84e-03 & 6.99e-02 & 1.75e-02\\
 & TXS0853+211 & 0.25 & 2.26e-03 & 4.11e-02 & 1.03e-02\\
 & PKS1000+26 & 0.25 & 2.22e-03 & 4.04e-02 & 1.01e-02\\
 & NVSSJ101808+190614 & 0.24 & 2.18e-03 & 3.96e-02 & 9.90e-03\\
 & MG1J101810+1903 & 0.24 & 2.18e-03 & 3.96e-02 & 9.90e-03\\
 & RXJ0908.9+2311 & 0.21 & 1.86e-03 & 3.39e-02 & 8.48e-03\\
 & B20922+31B & 0.08 & 7.49e-04 & 1.36e-02 & 3.41e-03\\
\hline
& & \textbf{23.23} & \textbf{2.1e-01} &\textbf{3.8}  & \textbf{9.4e-01} \\
\hline
30 & PKS0637-75 & 5.75 & 5.12e-02 & 1.03e+00 & 2.59e-01\\
 & PMNJ0251-8226 & 4.21 & 3.75e-02 & 7.57e-01 & 1.89e-01\\
 & SUMSSJ054923-874001 & 3.89 & 3.46e-02 & 6.99e-01 & 1.75e-01\\
 & PKS0736-770 & 1.45 & 1.29e-02 & 2.61e-01 & 6.53e-02\\
 & PMNJ0810-7530 & 1.36 & 1.21e-02 & 2.44e-01 & 6.11e-02\\
 & PKS0541-834 & 0.95 & 8.43e-03 & 1.70e-01 & 4.26e-02\\
 & PKS1029-85 & 0.82 & 7.28e-03 & 1.47e-01 & 3.68e-02\\
 & SUMSSJ074220-813139 & 0.00 & 0.00e+00 & 0.00e+00 & 0.00e+00\\
\hline
& & \textbf{18.43} & \textbf{1.6e-01} &\textbf{3.3}  & \textbf{8.3e-01} \\
\hline
33 & NVSSJ195547+021514 & 5.14 & 4.57e-02 & 1.20e+00 & 2.99e-01\\
 & RXJ1931.1+0937 & 2.72 & 2.42e-02 & 6.35e-01 & 1.59e-01\\
 & 1RXSJ194246.3+103339 & 1.69 & 1.51e-02 & 3.94e-01 & 9.86e-02\\
 & NVSSJ201431+064849 & 1.60 & 1.42e-02 & 3.73e-01 & 9.32e-02\\
 & TXS1923+123 & 0.94 & 8.33e-03 & 2.18e-01 & 5.45e-02\\
 & 1RXSJ193320.3+072616 & 0.75 & 6.63e-03 & 1.74e-01 & 4.34e-02\\
 & 87GB195252.4+135009 & 0.69 & 6.11e-03 & 1.60e-01 & 4.00e-02\\
 & NVSSJ190836-012642 & 0.45 & 3.99e-03 & 1.04e-01 & 2.61e-02\\
 & PMNJ2014-0047 & 0.36 & 3.22e-03 & 8.44e-02 & 2.11e-02\\
 & 87GB194635.4+130713 & 0.30 & 2.67e-03 & 7.00e-02 & 1.75e-02\\
 & 87GB201926.8+061922 & 0.24 & 2.16e-03 & 5.66e-02 & 1.41e-02\\
\hline
& & \textbf{14.87} & \textbf{1.3e-01} &\textbf{3.5}  & \textbf{8.7e-01} \\
\hline
38& NONE & 0.0 & 0.0 & 0.0\\
\hline
& & \textbf{0.00} & \textbf{0.0e+00} &\textbf{0.0}  & \textbf{0.0e+00} \\
\hline
\end{tabular}
\end{table}
\begin{table}[b]\centering\fontsize{5.5}{5.7}\selectfont
\addtocounter{table}{-1}
\caption{contd.}
\begin{tabular}{llllll}

IC & Source & \#$\nu$ & \#$\nu_{\mathrm{emp.}}$ & \#$\nu_{\mathrm{scale, int.}}$ & \#$\nu_{\mathrm{f}}$\\
\hline
39 & PKS0727-11 & 16.79 & 1.49e-01 & 3.10e+00 & 7.76e-01\\
 & TXS0637-128 & 5.72 & 5.09e-02 & 1.06e+00 & 2.64e-01\\
 & PKS0646-306 & 3.97 & 3.54e-02 & 7.35e-01 & 1.84e-01\\
 & PMNJ0608-1520 & 3.86 & 3.43e-02 & 7.13e-01 & 1.78e-01\\
 & PKS0648-16 & 3.17 & 2.82e-02 & 5.86e-01 & 1.46e-01\\
 & PKS0627-199 & 3.06 & 2.72e-02 & 5.66e-01 & 1.41e-01\\
 & TXS0628-240 & 2.61 & 2.32e-02 & 4.82e-01 & 1.20e-01\\
 & TXS0745-165 & 2.01 & 1.79e-02 & 3.71e-01 & 9.28e-02\\
 & TXS0700-197 & 1.86 & 1.65e-02 & 3.44e-01 & 8.59e-02\\
 & NVSSJ062753-152003 & 1.78 & 1.59e-02 & 3.30e-01 & 8.24e-02\\
 & PMNJ0634-2335 & 1.61 & 1.43e-02 & 2.98e-01 & 7.44e-02\\
 & TXS0616-116 & 1.58 & 1.41e-02 & 2.93e-01 & 7.32e-02\\
 & 1RXSJ064444.2-285120 & 1.33 & 1.18e-02 & 2.46e-01 & 6.15e-02\\
 & PMNJ0618-2426 & 1.29 & 1.15e-02 & 2.39e-01 & 5.97e-02\\
 & 1RXSJ072259.5-073131 & 1.15 & 1.03e-02 & 2.13e-01 & 5.33e-02\\
 & TXS0646-176 & 1.14 & 1.02e-02 & 2.11e-01 & 5.27e-02\\
 & TXS0752-116 & 0.97 & 8.64e-03 & 1.80e-01 & 4.49e-02\\
 & PMNJ0622-2605 & 0.84 & 7.45e-03 & 1.55e-01 & 3.87e-02\\
 & CRATESJ061733.67-171522.8 & 0.65 & 5.77e-03 & 1.20e-01 & 3.00e-02\\
 & TXS0728-054 & 0.62 & 5.54e-03 & 1.15e-01 & 2.88e-02\\
 & 1RXSJ064933.8-313914 & 0.45 & 3.98e-03 & 8.27e-02 & 2.07e-02\\
\hline
& & \textbf{56.47} & \textbf{5.0e-01} &\textbf{10.4}  & \textbf{2.6e+00} \\
\hline
40 & MRC1036-529 & 7.38 & 6.57e-02 & 1.29e+00 & 3.22e-01\\
 & PMNJ0845-5458 & 3.89 & 3.47e-02 & 6.80e-01 & 1.70e-01\\
 & PMNJ0852-5755 & 3.85 & 3.42e-02 & 6.72e-01 & 1.68e-01\\
 & (PMNJ1006-5018) & 3.71 & 3.30e-02 & 6.48e-01 & 1.62e-01\\
 & PKS0903-57 & 3.31 & 2.95e-02 & 5.79e-01 & 1.45e-01\\
 & PKS0920-39 & 2.41 & 2.15e-02 & 4.21e-01 & 1.05e-01\\
 & PMNJ1022-4232 & 1.85 & 1.64e-02 & 3.23e-01 & 8.06e-02\\
 & PMNJ1014-4508 & 1.22 & 1.09e-02 & 2.13e-01 & 5.34e-02\\
 & RXJ1023.9-4336 & 0.79 & 7.02e-03 & 1.38e-01 & 3.45e-02\\
\hline
& & \textbf{28.41} & \textbf{2.5e-01} &\textbf{5.0}  & \textbf{1.2e+00} \\
\hline
45& NONE & 0.0 & 0.0 & 0.0\\
\hline
& & \textbf{0.00} & \textbf{0.0e+00} &\textbf{0.0}  & \textbf{0.0e+00} \\
\hline
46 & PKS1004-217 & 3.00 & 2.67e-02 & 5.26e-01 & 1.31e-01\\
 & PMNJ1008-2912 & 1.75 & 1.55e-02 & 3.06e-01 & 7.66e-02\\
 & 1RXSJ102658.5-174905 & 1.18 & 1.05e-02 & 2.07e-01 & 5.18e-02\\
 & NVSSJ103040-203032 & 0.88 & 7.80e-03 & 1.54e-01 & 3.84e-02\\
 & TXS0956-244 & 0.52 & 4.64e-03 & 9.15e-02 & 2.29e-02\\
 & TXS0936-173 & 0.50 & 4.47e-03 & 8.81e-02 & 2.20e-02\\
 & 1RXSJ094709.2-254056 & 0.37 & 3.31e-03 & 6.52e-02 & 1.63e-02\\
\hline
& & \textbf{8.20} & \textbf{7.3e-02} &\textbf{1.4}  & \textbf{3.6e-01} \\
\hline
48 & PKSB1424-328 & 3.47 & 3.09e-02 & 6.63e-01 & 1.66e-01\\
 & PMNJ1347-3750 & 2.08 & 1.85e-02 & 3.97e-01 & 9.93e-02\\
 & 1RXSJ144037.4-384658 & 1.95 & 1.74e-02 & 3.73e-01 & 9.34e-02\\
 & PKS1348-289 & 1.61 & 1.43e-02 & 3.07e-01 & 7.69e-02\\
 & NVSSJ134543-335643 & 1.49 & 1.33e-02 & 2.85e-01 & 7.12e-02\\
 & SUMSSJ135625-402820 & 1.46 & 1.30e-02 & 2.80e-01 & 7.00e-02\\
 & PKS1341-366 & 1.04 & 9.27e-03 & 1.99e-01 & 4.97e-02\\
 & PMNJ1359-3746 & 0.72 & 6.40e-03 & 1.37e-01 & 3.43e-02\\
 & NVSSJ134706-295840 & 0.64 & 5.70e-03 & 1.22e-01 & 3.06e-02\\
 & NVSSJ142750-321515 & 0.22 & 1.99e-03 & 4.26e-02 & 1.07e-02\\
\hline
& & \textbf{14.69} & \textbf{1.3e-01} &\textbf{2.8}  & \textbf{7.0e-01} \\
\hline
52 & PMNJ1648-4826 & 29.81 & 2.65e-01 & 5.23e+00 & 1.31e+00\\
 & PMNJ1650-5044 & 7.45 & 6.63e-02 & 1.31e+00 & 3.27e-01\\
 & PMNJ1717-5155 & 7.23 & 6.43e-02 & 1.27e+00 & 3.17e-01\\
 & AT20GJ164513-575122 & 7.05 & 6.28e-02 & 1.24e+00 & 3.10e-01\\
 & PMNJ1711-5028 & 6.12 & 5.44e-02 & 1.07e+00 & 2.69e-01\\
 & MRC1613-586 & 5.77 & 5.14e-02 & 1.01e+00 & 2.53e-01\\
 & PMNJ1723-5614 & 1.75 & 1.56e-02 & 3.07e-01 & 7.67e-02\\
\hline
& & \textbf{65.18} & \textbf{5.8e-01} &\textbf{11.4}  & \textbf{2.9e+00} \\
\hline
\hline
All & Total &492.50 & 4.38  & 96.52 & 24.13 \\
\hline
\end{tabular}
\end{table}

\clearpage

\onecolumn
\section{High-energy spectral energy distributions}
This appendix shows the SEDs of all sources in the sample. [This
  section is only available in the journal version.]%

\end{appendix}

\end{document}